\RequirePackage{ifpdf}
\documentclass{PoS}
\usepackage{collref}
\title{Glueballs from the Lattice}

\ShortTitle{Glueballs from the Lattice}

\author{\speaker{Biagio Lucini}
         \thanks{In memory of Pierre van Baal, who greatly
           contributed to Lattice Gauge Theories.}\\
        Physics Department, Swansea University, Singleton Park, Swansea SA2 8PP, UK\\
        E-mail: \email{B.Lucini@swansea.ac.uk}}

\abstract{
Recent numerical calculations of the glueball spectrum in QCD, in
SU($N$) Yang-Mills theory in the large-$N$ limit and in candidate theories
of strongly interacting dynamics beyond the standard 
model (in which the lowest-lying scalar plays the role of the Higgs boson)
are reviewed and their implications for our theoretical
understanding of glueballs in QCD-like theories and in strongly coupled
gauge theories with a (near-)conformal dynamics are discussed.
}

\FullConference{From quarks and gluons to hadronic matter: A bridge too far?\\
   2-6 September, 2013\\
    European Centre for Theoretical Studies in Nuclear Physics and Related Areas (ECT*), Villazzano, Trento (Italy)}

\begin{document}

\section{Introduction}
One of the implications of colour confinement in SU$(N$) Yang-Mills
theories is the existence of massive gauge-invariant colour singlet
states with purely gluonic content, the {\em glueballs}. The coupling
being of order one,  a quantitative
understanding of the glueball spectrum transcend perturbative
techniques and can only be achieved in an {\em ab-initio}
framework. Among SU($N$) gauge groups, phenomenologically, a special
role is played by SU(3), which is the gauge group of QCD. One expects that
glueball states also appear in QCD. However, to date, glueballs have eluded
any experimental attempt aimed at their identification in the QCD
spectrum (see e.g.~\cite{Mathieu:2008me} for a recent review of the
theoretical and experimental status of this field). The fundamental
reason for the lack of experimental evidence of glueballs in the spectrum
can be traced back to the fact that there is no quantum number that
distinguishes glueballs from isosinglet mesons in the same $J^{PC}$
channel. Hence, physical states could naturally be thought of as
admixtures of states that in idealised conditions one would call
glueballs and states that are naturally identified as isosinglet mesons. This
mixing can be  assumed to be the result of some off-diagonal effective
Hamiltonian that couples nearby states in the spectrum with the same
quantum numbers. If the mixing is small, then one could classify the
states that are coupled by this Hamiltonian under gluon-rich and 
quark-rich. However, it is a dynamical problem to establish the
strength of the mixing terms. 

As for other non-perturbative features of QCD, numerical calculations with Monte
Carlo methods in the theory discretised on a spacetime lattice
provide invaluable insights on the nature of glueballs~(see
e.g.~\cite{McNeile:2008sr} for a recent review and~\cite{Gregory:2012hu} and
references therein for more recent developments). Those calculations
should measure the effective couplings for the production
and annihilation of glueballs and of mesons (which will be related to
the diagonal elements of the effective Hamiltonian describing the
interactions of those states) as well as the couplings between
glueballs and mesons. Although progress in this direction has been
achieved, a calculation of this type is still unviable. Thus, most of current
lattice calculations assume that the mixing is small and compute
masses of mesons and glueballs separately, the hope being that the
real interacting spectrum will not look much different from this
simplified case. In addition, since glueball measurements often
demand a large statistics, which in turn requires a huge computational
effort, the glueball spectrum is often provided in the quenched
approximation, in this case amounting to studying the problem in
the SU(3) Yang-Mills theory. The reference calculation in this
approximation is still provided by~\cite{Morningstar:1999rf}, with
some subsequent work~\cite{Meyer:2004jc,Chen:2005mg} confirming the
main results and providing some extra insights.
The quenched approximation, which at first sight looks quite drastic, can in
fact be formulated in a framework that provides a controlled
interpolation from QCD to SU(3) Yang-Mills: the 't Hooft large-$N$
limit~\cite{'tHooft:1973jz} (see ~\cite{Coleman:1980nk} for an early
pedagogical review of the main results).

In addition to the theoretical and experimental interest for glueballs
in QCD, more recently there has been an increasing interest in
glueball-like states in theories that like QCD confine in the
infrared and are asymptotically free in the ultraviolet, but unlike
QCD have an intermediate energy region in which the physics is
governed by an infrared fixed point in an enlarged parameter
space. These theories, normally referred as
{\em technicolour}, are said to be walking, since the coupling runs very
slow in this intermediate energy range\footnote{Although the coupling
  is not a physical quantity, the concept of running can be formalised
  in terms of spectral observables.}. Walking, and more in general
infrared conformal gauge theories, can provide a dynamical mechanism
of electroweak symmetry breaking. A full discussion of this mechanism
goes beyond the scope of this work, and we refer the interested reader
to the specialised literature (e.g.~\cite{Andersen:2011yj} and
references therein). Our interest here is 
in the fact that in this framework the lowest-lying isosinglet scalar
plays a special role, since it should be identified with the recently
discovered Higgs boson. Once again, the existence of such a light
scalar is a dynamical problem that can be addressed with lattice
calculations. 

In this contribution, we give a brief overview of recent lattice
results. This work does not aim at providing a comprehensive review, nor
does it pretend to be complete. The main purpose here is to discuss
some of the most recent results for glueball masses obtained from
lattice calculations and their possible implications for our
understanding of glueballs. A more detailed account
of the field is provided in the quoted references and in works cited
therein. The rest of this paper is structured as follows. In
Sect.~\ref{sect:2} we briefly review the lattice formulation of
SU($N$) gauge theories and of QCD, with a particular emphasis on
widely employed techniques to extract glueball masses. Sect.~\ref{sect:3}
discusses numerical results in SU($N$) Yang-Mills and their relevance
for QCD, providing also an explicit example of a QCD
calculation. Calculations of the glueball spectrum in theories
potentially relevant for explaining dynamically electroweak symmetry
breaking are reported in Sect.~\ref{sect:4}. Finally,
Sect.~\ref{sect:5}  summarises the main lessons learnt from those
calculations and points out some of the current and future directions
of lattice calculations of glueball properties.

\section{Extracting glueball masses from lattice calculations}
\label{sect:2}
The lattice is the most reliable method for obtaining {\em ab-initio}
quantitative predictions in the non-perturbative regime of QCD. The
calculation starts with the QCD action regularised on a
four-dimensional spacetime grid in such a way that gauge invariance is
preserved. In the pure gauge sector, the simplest choice is the Wilson action:
\begin{equation}
S_w = \beta \sum_{i, \mu > \nu} \left( 1 - \frac{1}{N} \ {\cal R}e
  \mbox{Tr} \  
U_{\mu \nu}(i) \right) \ .
\end{equation}
In the previous equation, $i$ labels a point in the discretised
spacetime, $\mu$ and $\nu$ are two directions, and the sum extends
over all points and directions. $ {\cal R}e \mbox{Tr} \ U_{\mu \nu}$ is the
real part of the trace of the {\em plaquette} $U_{\mu \nu}$, with the
latter being the path ordered product of the oriented link variables
$U_{\mu}(i)$ (with reverse orientation corresponding to Hermitian
conjugation) around an elementary square of the lattice.
Positive orientations of the links connecting nearest neighbour sites
follow the direction of the versors $\hat{\mu}$. A link variable
$U_{\mu}(i)$  is an element of SU(3) (or of the gauge group, in the
most general case),  related to the continuum field $A_{\mu}$ by
\begin{equation}
U_{\mu}(i) = {\mbox P} \exp \left(i g a \int_i ^{i + \hat{\mu}}  A_{\mu}(x)
  \mbox{d} x \right) \ ,
\end{equation}
with the integral taken from point $i$ to its nearest neighbour $i +
\hat{\mu}$ along the direction $\hat{\mu}$ and $a$ being the lattice
spacing. The lattice coupling 
$\beta$ is related to the continuum (bare) coupling $g$ by $\beta = 2
N / g^2$, with $N$ the number of colours (i.e. $N = 3$ for QCD). Gauge
transformations are represented by functions $G(i)$ defined on the
lattice points and taking values in the gauge group. They act on link
variables by sending $U_{\mu}(i)$ to $G(i)^{\dag} U_{\mu}(i) G(i +
\hat{\mu})$, where the $\dag$ symbol indicates Hermitian
conjugation. It is straightforward to see that with these definitions
the action $S_w$ is gauge invariant. It is also easy to see that, in
the limit $a \to 0$, at the lowest order in $a$, $S_w$ reproduces the
Yang-Mills continuum action. Using asymptotic freedom, it is
possible to show that quantum fluctuations do not spoil the
correctness of the continuum limit. 

For fermions, on the lattice one can not preserve at the same time
chiral symmetry and ultralocality of the action without generating
spurious copies called {\em doublers}. Hence, if we want to
have an action coupling a finite number of neighbour points, we have to
give up either the possibility of having a generic number of
flavours or exact chiral symmetry. Chirally symmetric fermionic actions with an
arbitrary number of flavours can also be formulated, at the expenses
of coupling all the sites (or, equivalently, of adding
an extra dimension). It is worth stressing that (as one would
expect) all lattice fermionic actions reproduce the same continuum
physics, but some of them can be used more proficiently than others in
particular physical situations. Without going further into specialistic technical
details, the actions describing the fermion-gauge interactions we are
interested in can be written as quadratic forms in terms of the
fermion fields:
\begin{equation}
S_f = \overline{\psi}_i M_{ij} \psi_j \ ,
\end{equation}
where $i$ and $j$ are Euclidean indices, with all other indices kept
implicit. $M$ embodies the spinorial structure and depends linearly on
the gauge fields $U_{\mu}$. The path integral of the full theory is
\begin{equation}
\label{eq:zqcd}
Z = \int {\cal D} U_{\mu} \left(\mbox{Det}\ M \right)^{N_f} e^{- S_w} \ ,
\end{equation}
where $N_f$ is the number of flavours, $ {\cal D} U_{\mu}$ is the
path integral measure of all links (which on the lattice is well
defined and corresponds to a product of Haar measures) and the
determinant of $M$ comes after the Grasmann integral over the fermions.  
Eq.~(\ref{eq:zqcd}) is amenable to Monte Carlo numerical
simulations. State of the art numerical results require modern supercomputers.

Using the path integral~(\ref{eq:zqcd}), the vacuum
expectation value of an operator ${\cal O}$ depending only on the link
variables is obtained as
\begin{equation}
\label{eq:vev}
\langle {\cal O} \rangle = \frac{1}{Z} \int {\cal D} U_{\mu}
\left(\mbox{Det}\ M \right)^{N_f} {\cal O} e^{- S_w} \ .
\end{equation}
For glueball mass calculations,  ${\cal O}$ tales the form of a
correlator of an appropriate observable $\overline{O}$ with the quantum numbers
of the state of interest. If $\overline{O}$ depends on the point $i =
(\vec{x},t)$, where we have conventionally separated the Euclidean
time coordinate $t$ and the space coordinates $\vec{x}$, we define
$O(t)$, the zero-momentum operator at timeslice $t$, as
\begin{equation}
O(t) = \frac{1}{V_s} \sum_{\vec{x}} \overline{O}(x,t) \ ,
\end{equation}
with $V_s$ the spatial volume. The correlator of interest is $C(\tau)
= \langle O^{\dag}(t_0) O(t_0 + \tau) \rangle$, which, by time
translational invariance, is a function of $\tau$ only. Using the
spectral representation,
\begin{equation}
C(\tau) = \int {\cal C}(E) e^{- E \tau} \mbox{d}E \ .
\end{equation}
At the lowest end of the spectrum, the spectral function
${\cal C}(E)$ is a discrete sum over Hamiltonian eigenstates with eigenvalues
corresponding to physical masses of states contributing to
$C(\tau)$. In the limit $\tau \to \infty$, only the lowest state with
the quantum number of $\overline{O}(x,t)$ contributes to ${\cal C}(\tau)$. 
If $\overline{O}(x,t)$ is taken to transform under an
irreducible representation of the rotational group, its energy will be
the lowest mass $m_J$ in the channel with spin corresponding to that
irreducible representation:
\begin{equation}
\label{eq:effmass}
m_J = - \lim_{\tau \to \infty} \frac{C(\tau)}{\tau} \ .
\end{equation}
Since the lattice has a finite extent, the previous formula
can be reliably used to extract the mass of glueballs as
long as $\overline{O}$ is chosen in such a way that the contribution of higher
excitations is minimised. An optimal choice of $\overline{O}$ can be
performed only if the Hamiltonian eigenstates are known, but these are
exactly what we are after. However, we can approximate these
eigenstates using a variational method. In practice, we use a
large set of operators $\overline{O}_i$ by considering traced path ordered
products of link variables over closed contours of various shapes and
sizes. Each fixed contour is transformed under the rotational group
and the results are combined in such a way that irreducible
representations are constructed. Using paths transformed under the
appropriate symmetries, we build combinations that have also well
defined parity and charge conjugations. 

The same base paths can give an increased basis of operators by
replacing the links with more extended objects. These new variables,
often referred to as {\em fat links}, are still defined on lattice
links, with a modification of the original link variables that
accounts also for contributions
coming from a neighbourhood of each link. The new links transform
under gauge transformations as the original ones. The physical idea
behind this procedure is that hadrons are extended objects,
while the links are ultraviolet quantities. If we want to access the
properties of physical particles we need to use variables that have support
on the typical physical length. In practice, one replaces original
variables with blocked variables, as in a block-spin transformation in
statistical mechanics. These blocking transformations can be combined
and iterated. There are various ways of implementing this idea. The
results we review in this work follow Refs.~\cite{Lucini:2004my,Lucini:2004eq}.

Having built a set of loops with well defined $J^{PC}$ transformation
laws, we consider all possible correlators $C_{ij}(\tau) = \langle
O^{\dag}_i(t_0) O_j(t_0 + \tau) \rangle$, where each operator is the trace of
the ordered product  over some fixed base loop of either the original
links or the link variables obtained after a given number of
iterations of the blocking transformations. The
correlation matrix $C_{ij}(\tau)$ (normalised so that $C_{ij}(0)$ is the
identity) is then diagonalised for a fixed
$\tau$ (generally, $\tau = 1$; although a higher $\tau$ would be
desirable, rarely the statistics is sufficient to get an accurate
signal for $\tau > 1$). This procedure defines the eigenstates $\Phi_1(1),\
\dots, \Phi_M(1)$, where $M$ is the dimension of $C$. For simplicity,
we consider a case in which the eigenvalues are non-degenerate, the
more general case only posing technical complications that could be
dealt with quite easily. The eigenvectors are assumed to be ordered
from the one corresponding to the higher to the one corresponding to
the lower eigenvalue. Applying  Eq.~(\ref{eq:effmass}) to $\Phi_k$,
one obtains a variational estimate of the mass for the excitation $k-1$
with quantum numbers $J^{PC}$. 

In practice, things are more complicated. For a start, only a few low
states are below the onset of a continuum spectral density, which
starts with the multi-particle bound state of lowest possible energy. This
requires us to identify scattering states and to project them out from
our basis. Second, the spatial lattice has not the full rotational
symmetry, but has the symmetry of the invariance group of the
three-dimensional cube, the octahedral group, which has a finite number
of elements (24 excluding parity). Hence, one needs to classify the operators according
to the transformations of the octahedral group and then decompose the
latter according to the continuum spin contributions, with more than
one continuum spin contributing to each irreducible representation of
the octahedral group. Then, on a finite lattice, there are spurious
contributions to the glueball spectrum related to operators winding
around the periodic spatial directions. These contributions, called
{\em torelon states}, need to be removed from the spectrum. Finally,
one has to keep in mind that the calculation is variational. While
this is not a big problem for the lowest-lying states, the variational
estimate of higher excitations is less and less accurate, since the
number of operators in the variational basis decreases as we go higher
in mass. These issues have been known since the early days of
glueball calculations on the lattice (see
e.g.~\cite{Berg:1982kp,Michael:1988jr}). Various approaches have been 
developed over the years, and glueball calculations for the
ground state and a few excitations in some spin channels (e.g. the
$0^{++}$ 
and  the $2^{++}$) are already quite accurate, at least in the pure
Yang-Mills case. Specific calculations have targeted other quantum
states, providing a rather comprehensive set of results, as for
instance in
~\cite{Morningstar:1999rf,Meyer:2004jc,Chen:2005mg}. Recently, the
work~\cite{Lucini:2010nv} has appeared, in which the construction of the variational
basis is fully automatised. This method
allows us to perform a full calculation in all the channels, in
which we can identify scattering and torelon contributions and hence
remove them from the physical spectrum. This work -- to which we refer
for further details -- is the foundation for the main results
presented in the next section.

\section{From SU(3) to SU($\infty$) and back to QCD}
\label{sect:3}
For a long time, performing calculations at quark masses that are low
enough for  being phenomenologically relevant has been a big
practical problem
in lattice QCD. Recent advances in numerical algorithms and
computational power allow calculations to be done at a pseudoscalar mass
close to that of the physical $\pi$. However, this still poses a
computational challenge that limits the statistics. The number of
configurations that can be obtained are generally sufficient for
getting the meson spectrum at an acceptable accuracy level. However,
numerical calculations of glueball masses require higher statistics (at
least by a factor of ten) Monte Carlo simulations. The reason for the
different statistical requirements to get an error of the same order
of magnitude is well understood, and can be explained in terms of QCD
dynamics. While the field is progressing at a fast pace, still the
required computational  time is highly non-trivial to obtain.
Because of these high computational demands, it still makes sense to
resort to the so-called quenched approximation, in which the quarks
are treated as external sources. In glueball calculations of SU(3),
this amounts to studying the pure Yang-Mills theory. At this stage, it
is an open problem to establish whether and how these calculations
could be relevant for real-world QCD.  

\begin{figure} 
  \begin{center}
    \includegraphics[width=1.0\textwidth]{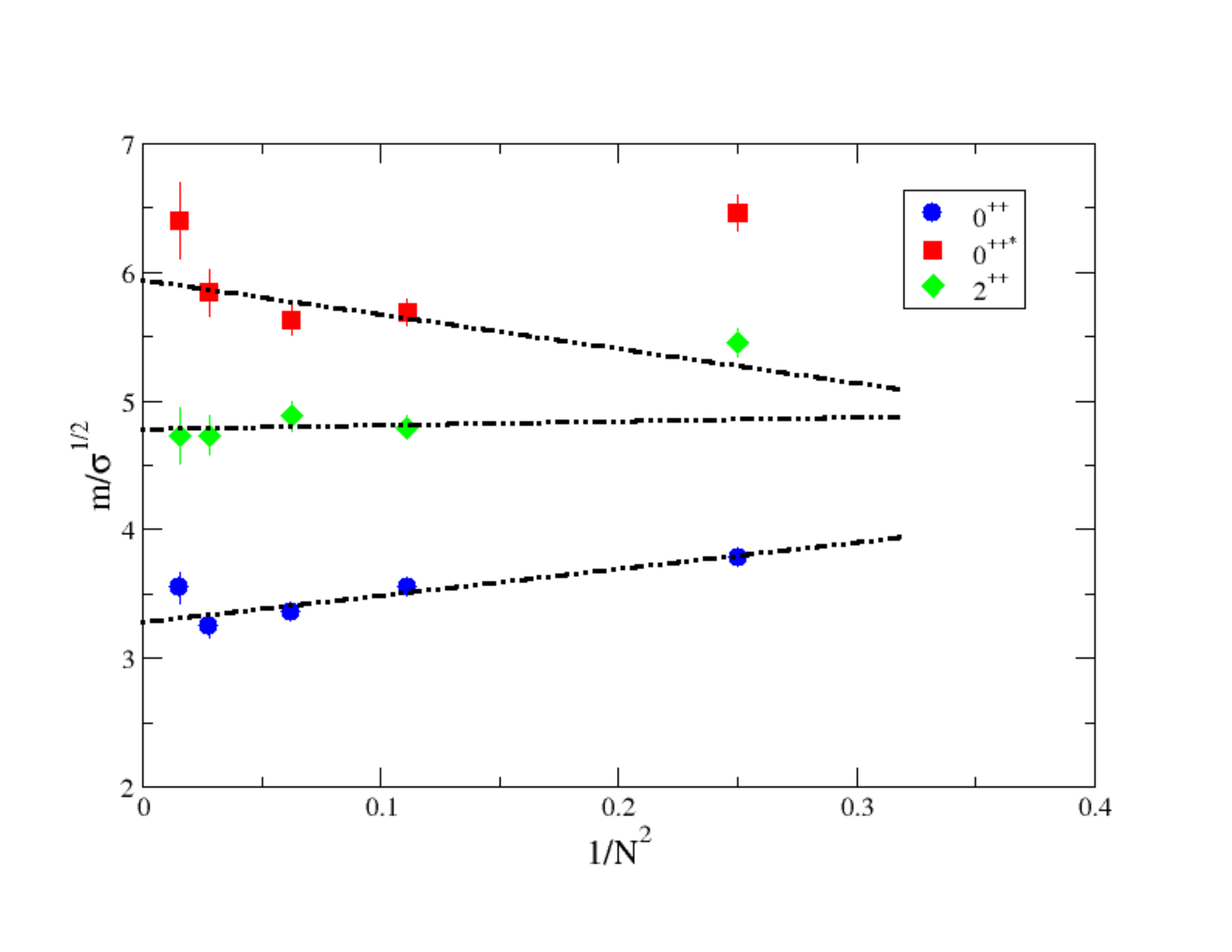} 
  \end{center}
  \vspace{-1cm}
  \caption{Extrapolations to $N \to \infty$ of the lowest-lying
    glueball masses.} 
  \vspace{-0.23cm}
  \label{fig:sun_gall}
\end{figure}
 
In fact, a natural framework to interpolate from SU(3) to QCD, and
hence to ascertain the importance of glueball calculations in pure
gauge for the full dynamical theory, is provided by the 't
Hooft large-$N$ limit.  In this context, pure SU(3)
gauge theory is seen as a particular case of the generic family of SU($N$) Yang-Mills theory. As
the number of colours $N$ is sent to infinity and the coupling of the
theory $g$ is sent to zero with the constraint that the 't Hoof
coupling $\lambda =g^2 N$ remains constant, at the perturbative level
the theory simplifies, and one can prove that observables can be
expressed in a power series in $1/N^2$. If the coefficients of this
series are of order one, one would expect the physics in SU(3) to
only differ from that of the large-$N$ extrapolated theory by a
quantity of the order of 10\% in each observable. 

Numerically, the determination of the large-$N$ limit of SU($N$) gauge theories
can be achieved  by studying the values of observables for various
SU($N$) group and assuming the validity of the diagrammatic argument
that suggests a power series
in$1/N^2$~\cite{Lucini:2001ej,Lucini:2001nv}.  For various
observables, this procedure suggests that at the level of accuracy of
a few percents the SU(3) value and the infinite-$N$ value can be related
using only the leading correction~(see~\cite{Lucini:2012gg,Lucini:2013qja} and
references therein). 

\begin{figure} 
  \begin{center}
    \includegraphics[width=0.9\textwidth]{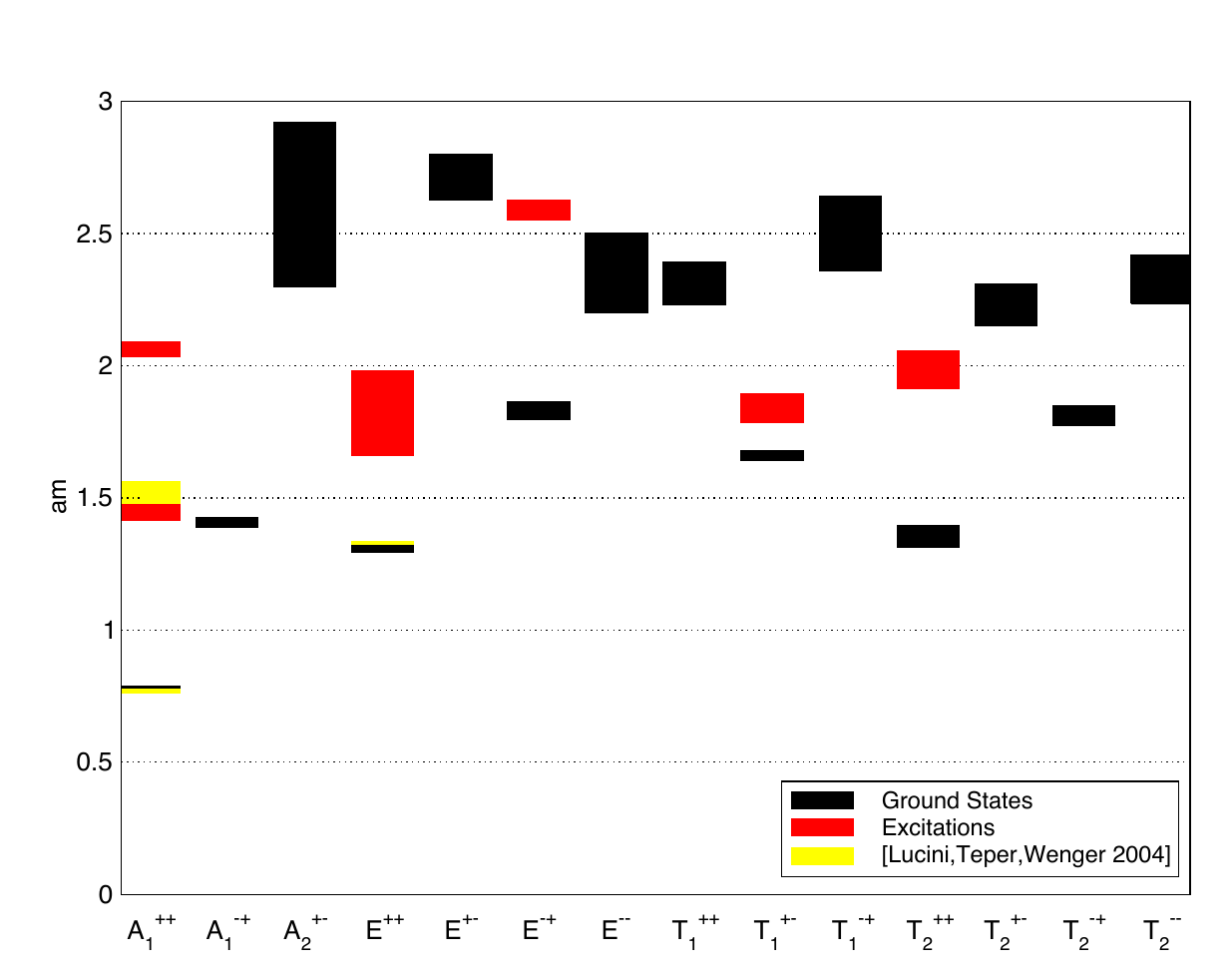} 
  \end{center}
  \caption{Glueball masses in various channels (ground states and some
    excitations) in the large-$N$ limit at fixed lattice spacing $a =
    1/(6 T_c)$, $T_c$ being the deconfinement temperature.} 
  \label{fig:glueball_spectrum}
\end{figure}

The most precise determination of the SU($N$) continuum limit of the
lowest-lying glueball masses has been provided
in~\cite{Lucini:2004my} (see Fig.~\ref{fig:sun_gall}). Using the square root of the string tension
$\sqrt{\sigma}$ to set the scale, the quoted results for the lowest mass in the
$0^{++}$ channel, the lowest mass in the $2^{++}$ channel and the
first $0^{++}$ excitation are respectively
\begin{eqnarray}
  \nonumber
  \frac{m_{0^{++}}}{\sqrt{\sigma}} &=& 3.28(8) + \frac{2.1(1.1)}{N^2}
\ , \\
  \frac{m_{2^{++}}}{\sqrt{\sigma}} &=& 4.78(14) + \frac{0.3(1.7)}{N^2}
  \ , \\
  \nonumber
  \frac{m_{0^{++*}}}{\sqrt{\sigma}} &=& 5.93(17) -\frac{2.7(2.0)}{N^2}
  \ .
\end{eqnarray}
The emerging picture is that, at level of accuracy of around 5-10\%,
glueballs are well described by their large-$N$ limit, the correction
coefficient being of order one. This is consistent with the expectation
of the large-$N$ series being convergent at $N = 3$. In this sense,
and following the fact that the corrections are of the order of
magnitude one would naively expect, one can say that SU(3) is close to
SU($\infty$). Hence, it makes sense to look at the physics of
glueballs in the simpler $N = \infty$ theory. A calculation
performed in~\cite{Lucini:2010nv} shows -- albeit at fixed lattice
spacing -- that similar conclusions seem also to be valid for all
glueball states currently accessible to numerical simulations (Fig.~\ref{fig:glueball_spectrum}).
This is good news for large-$N$ based approaches to glueballs and
QCD-like strong interactions, like for instance the gauge-string
duality (see e.g.~\cite{Elander:2013jqa} and references therein) or,
more recently, topological field theory~\cite{Bochicchio:2013sra}.    

\begin{figure} 
  \begin{center}
  \includegraphics[width=1.0\textwidth]{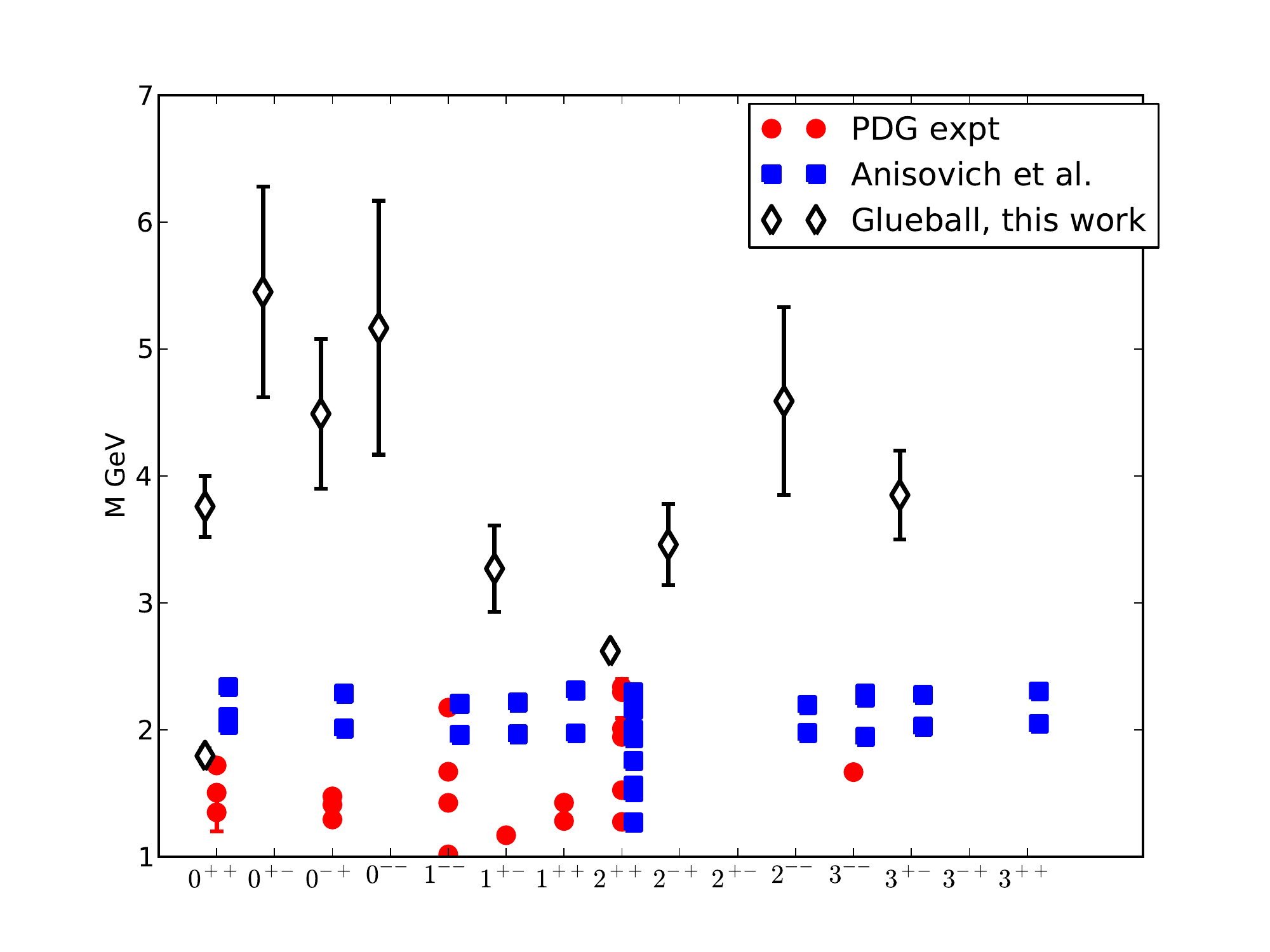} 
  \caption{Glueball masses in full QCD from a lattice
    calculation with dynamical fermions (open points) compared to experimental results, as
    described in~\cite{Gregory:2012hu}, from which the plot has been taken.}
  \end{center}
  \label{fig:glueball_sum}
\end{figure}

When dynamical fermions are considered, it can be shown that they
provide a leading contribution that is $1/N$-suppressed with respect
to the Yang-Mills leading term. Hence, if SU(3) is close to 
SU($\infty$) and the contribution given by quarks stays negligible
for $N =3$ (as recent lattice calculations seem to confirm~\cite{Bali:2013kia,Bali:2013fya}), then
glueball calculations in SU(3) gauge theory can still provide a useful guidance
to experiments. This provides a motivation to perform high precision
quenched calculations. 

Another relevant consequence of the large-$N$ limit is that mesons and
glueballs do not mix for infinite number of colours.  If SU(3) is
close to the large-$N$ limit, the suppression of mixing at large-$N$ can justify an
approach to glueballs that  neglects their mixing with mesons. Very
recently, high statistics numerical calculations of the glueball
spectrum in a full QCD setup have been performed
in~\cite{Richards:2010ck} and extended to include more spectral states
in~\cite{Gregory:2012hu}. The central results of the latter calculation are shown
in Fig.~\ref{fig:glueball_sum}, which compares the results of the
simulations (open points) with experimental states. Most of the
glueball states seem to be at quite high mass, in a region that will
be accessible to the PANDA experiment~\cite{Lutz:2009ff}. A current focus of lattice
calculations is to reduce both the statistical error and the systematic
error (e.g. by taking into account mixing with mesons and improving
spin identification). In particular, lattice calculations are
focusing on exotic quantum numbers (as for instance $2^{+-}$ and
$0^{-+}$), since these can not be explained by the simple quark model.
\section{Glueballs in near-conformal gauge theories}
\label{sect:4}
QCD provides an exemplar realisation of confinement and chiral
symmetry breaking with the physics determined only by one mass
scale. However, it has been conjectured for a long time that the
particular realisation of QCD is not the only possibility for having
confinement in the deep infrared in asymptotically free non-Abelian
gauge theories. Another possibility is that the theory be approximately infrared
conformal. Infrared conformal theories are realised for a number
of fermions close to the point at which the system ceases to be
asymptotically free. The range of number of flavours for which the
theory is infrared conformal is called {\em conformal window}. For
theories just below the onset of the conformal window the infrared
fixed point is only approximate, and for a wide range of energies 
observables have a mild dependence on the energy. Theories in this
regime are said to be {\em walking} or {\em near-conformal}. A walking or a
conformal novel interaction could provide a dynamical explanation of
electroweak symmetry breaking.

\begin{figure} 
  \begin{center}
    \includegraphics[width=.85\textwidth]{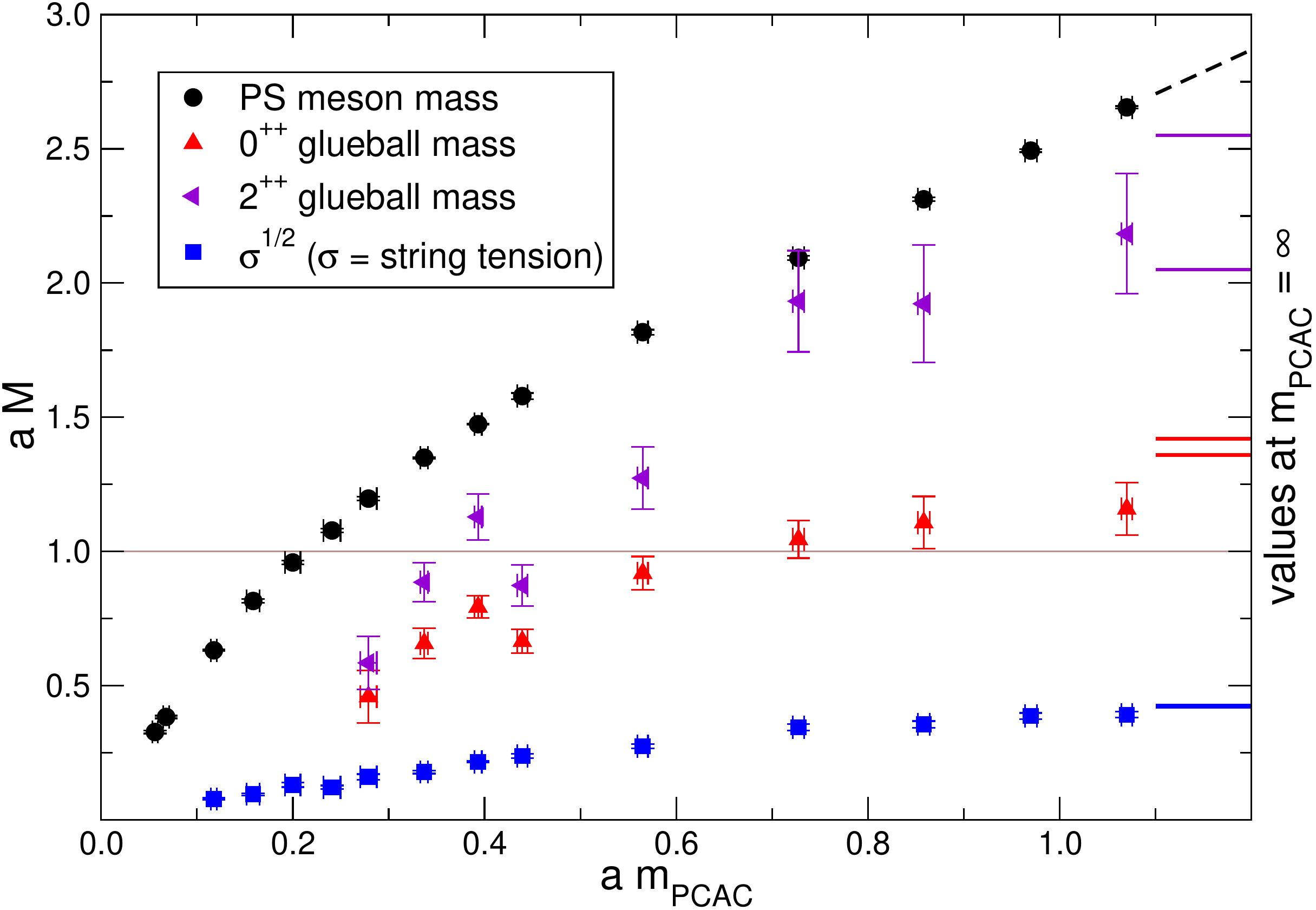} 
    \end{center}
  \caption{The lowest-lying states in the spectrum of SU(2) with two
    adjoint Dirac quarks~\cite{DelDebbio:2009fd,DelDebbio:2010hx}, which has an infrared fixed point.} 
  \label{fig:all}
\end{figure}

 As with other strong interactions,
the lattice approach can provide essential information on the physical
properties of those theories. Because of their potential relevance for
phenomenology beyond the standard model, recently those theories have been
widely studied with numerical simulations. One of the very first
objectives has been the determination of the conformal window, which
allows to pin down some candidate models for phenomenology. As a
fundamental step, one must be able to disentangle features of
(near-)conformality from a typical confining behaviour when the
fermions have a finite (albeit small) mass. One of the
signatures of a conformal gauge theory is a spectrum whose mass ratios do
not depend on the fermion mass, possibly presenting an inverted mass
hierarchy (i.e., unlike in QCD, glueball masses can be lower than the pseudoscalar
meson
mass~\cite{DelDebbio:2009fd,DelDebbio:2010hx,Aoki:2013zsa,Athenodorou:2013eaa},
see e.g. Fig.~\ref{fig:all}).
This feature is particularly important in the light of the recent
discovery of the Higgs particle, which in this framework will be
identified with a scalar that is anomalously light near a point in
which scale invariance gets restored. Although we have not seen yet this
mechanism at work in gauge theories that can have phenomenological
relevance, there are examples in gauge-string
duality~\cite{Elander:2012yh} and in lattice gauge theories in a
strongly coupled regime far from the continuum
limit~\cite{Lucini:2013wsa}. In particular, lattice gauge theories at
strong coupling can provide useful insights on the structure of
conformal or nearly-conformal gauge
theories~\cite{deForcrand:2012vh,Tomboulis:2012nr}.

\section{Conclusions}
\label{sect:5}
The experimental identification of glueballs is one of the most urgent
open problems in QCD. Thanks to recent progress, lattice calculations
are getting closer and closer to providing firm theoretical
predictions for glueball masses in all possible channels. Further progress will include a
quantification of the mixing with mesons, a systematic study of
states with exotic quantum numbers and a more accurate reconstruction
of the continuum spin from the lattice data.  More recently, glueballs (or
better, $0^{++}$ isosinglet states with a non-negligible contribution
coming from the pure glue dynamics) are starting to play a key role in
understanding the dynamics of strongly-coupled conformal or
near-conformal gauge theories and in their phenomenological
signatures, with a light glueball spectrum (originally identified
for one of those models in~\cite{DelDebbio:2009fd,DelDebbio:2010hx})
now starting to emerge as a key signature of those
theories~\cite{Aoki:2013zsa,Athenodorou:2013eaa}. Whether electroweak
symmetry breaking is described by a gauge theory in the near
conformal-phase and the Higgs particle should be seen as a {\em
  glueball} of a new strong interaction is current the subject of
a fervid theoretical and phenomenological investigation, with the
lattice proving to be once more an invaluable quantitative tool.

\acknowledgments
The results reviewed here have been obtained in various collaborations
with L. Del Debbio, E. Gregory, A. Irving,
C. McNeile, A. Patella, C. Pica. A. Rago, E. Rinaldi, M. Teper and U. Wenger. A special
thanks goes to Craig McNeile, Antonio Rago and Enrico Rinaldi for their constant and
crucial involvement in more recent glueball calculations. This work has  been partially supported by the Royal
Society grant UF09003 and the STFC grant ST/G000506/1. We also thank
High Perfomance Computing Wales for computational resources.

\bibliographystyle{JHEP}
\bibliography{lucini_qcdtnt}

\end{document}